\documentstyle[prl,aps,psfig]{revtex}
\draft
\begin{document}
\title{Quantum Memory for Light}
\author{A.~E.~Kozhekin, K.~M{\o}lmer and E.~Polzik}
\address{Institute of Physics and Astronomy,
University of Aarhus,
Ny Munkegade, DK-8000 Aarhus C, Denmark}
\date{\today}

\maketitle
\begin{abstract}
 We propose an efficient method for mapping and storage of a quantum
 state of propagating light in atoms. The quantum state of the light
 pulse is stored in two sublevels of the ground state of a macroscopic
 atomic ensemble by activating a synchronized Raman coupling between
 the light and atoms.  We discuss applications of the proposal in
 quantum information processing and in atomic clocks operating beyond
 quantum limits of accuracy. The possibility of transferring the
 atomic state back on light via teleportation is also discussed.
\end{abstract}

\pacs{42.50.Lc, 03.67.-a, 42.50.Dv, 42.50.Ct, 06.30.Ft}

%
%
Light is an ideal carrier of quantum information, but photons are
difficult to store for a long time. In order to implement a storage
device for quantum information transmitted as a light signal, it is
necessary to faithfully map the quantum state of the light pulse onto
a medium with low dissipation, allowing for storage of this quantum
state. Depending on the particular application of the memory, the next
step may be either a (delayed) measurement projecting the state onto a
certain basis, or further processing of the stored quantum state,
e.g., after a read-out via the teleportation process. The delayed
projection measurement is relevant for the security of various quantum
cryptography and bit commitment schemes \cite{Bras}. The
teleportation read-out is relevant for full scale quantum computing.

In this Letter we propose a method that enables quantum state transfer
between propagating light and atoms with an efficiency up to 100\% for
certain classes of quantum states. The long term storage of these
quantum states is achieved by utilizing atomic ground states. In the
end of the paper we propose an atom-back-to-light teleportation scheme
as a read-out method for our quantum memory.

\begin{figure}
{\centerline{
\psfig{file=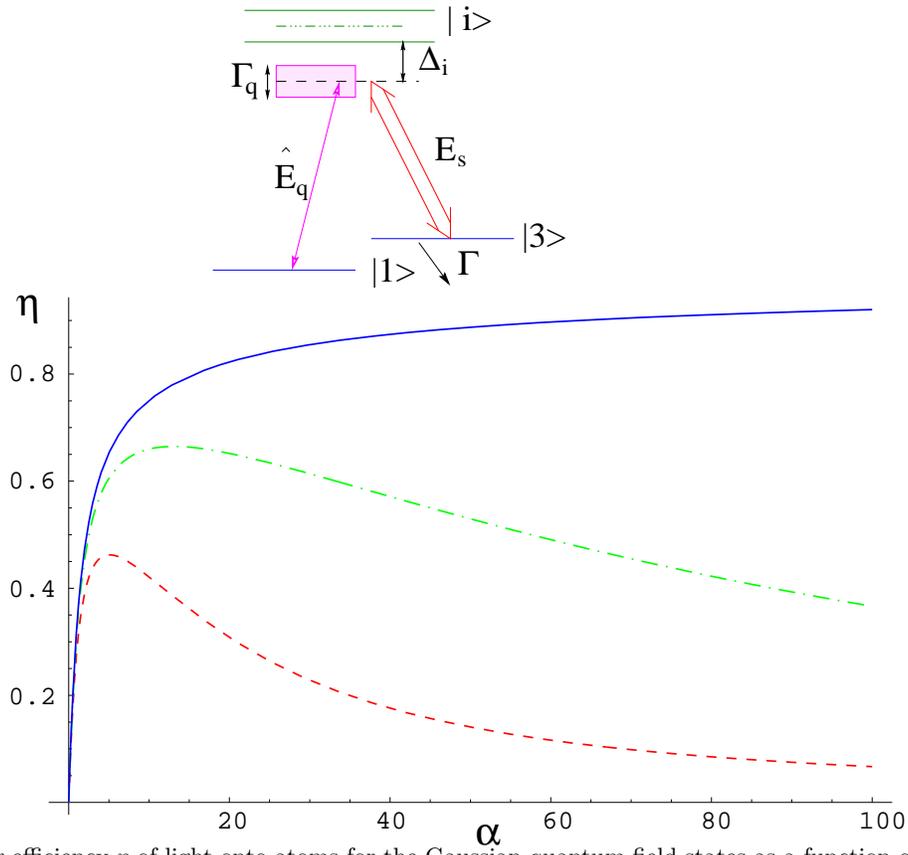,width=12cm}
}}
\caption{\label{fig:var} Mapping efficiency $\eta$ of light onto atoms
for the Gaussian quantum field states as a function of optical depth:
solid line -- infinite band squeezing, dash-dotted line
$\Gamma_q/\Gamma=50$, dashed line $\Gamma_q/\Gamma=10$.  Inset --
Schematic representation of atomic levels.}
\end{figure}

We consider the stimulated Raman absorption of propagating quantum
light by a cloud of $\Lambda$ atoms. As shown in the inset of
Fig.\ref{fig:var}, the weak quantum field and the strong classical
field are both detuned from the upper intermediate atomic state(s) by
$\Delta$ which is much greater than the strong field Rabi frequency
$\Omega_{s}$, the width of an upper level $\gamma_{i}$ and the
spectral width of the quantum light $\Gamma_{q}$. The Raman
interaction ``maps'' the non-classical features of the quantum field
onto the coherence of the lower atomic doublet, distributed over the
atomic cloud.

In our analysis we eliminate the excited intermediate states, and we
treat the atoms by an effective two-level approximation. We start with
the quantum Maxwell-Bloch equations in the lowest order for the slowly
varying operator $\hat{Q}$: $\hat{Q}=\hat{\sigma _{31}}
e^{-i(\omega_{q} - \omega_{s})t +i (k_{q}-k_{s})z}$ (it will be
assumed, that $(k_{q}-k_{s}) L \ll 1$, where $L$ is the length of the
atomic cloud, $z$ is the propagation direction, and $\omega_{q,s}$ and
$k_{q,s}$ are frequencies and wavevectors of ``quantum'' and
``strong'' fields respectively) \cite{Raym81,Raym85}
\begin{mathletters}
\begin{eqnarray}
&& \frac{d}{dt}\hat{Q}(z,t) =-i\kappa_{1}^{\ast} \hat{E}_{q}(z,t)
E_{s}^{\ast} (z,t) - \Gamma \hat{Q}(z,t) + \hat{F}(z,t)  \label{Bloch} \\
&& \left( \frac{\partial}{\partial z} + \frac{1}{c} \frac{\partial}{\partial t}
\right) \hat{E}_{q}(z,t) = -i \kappa_{2} \hat{Q}(z,t) E_{s}(z,t)
\label{Max}
\end{eqnarray}
\end{mathletters}
$\Gamma$ is the dephasing rate of the $1\leftrightarrow 3$ coherence
which also includes the strong field power broadening $\Gamma_{s}
\simeq \omega^{3} \hbar \kappa_{1}^{2} |E_{s}|^{2}/(3c^{3})$ due to
spontaneous Raman scattering \cite{Raym81}, $\hat{F}(z,t)$ is the
associated quantum Langevin force with correlation function $\langle
\hat{F^{\ast}}(z,t) \hat{F}(z^{\prime}, t^{\prime })\rangle =2\Gamma
/n \delta (z-z^{\prime })\delta (t-t^{\prime })$, and $\kappa_{1} =
\sum_{i} \mu_{1i} \mu_{3i}/(\hbar^{2} \Delta_{i})$, $\kappa_{2} = 2\pi
n\hbar \omega \kappa_{1}/c$, where $\mu_{ji}$ are dipole moments of
the atomic transitions and $n$ is the density of the atoms. A
one-dimensional wave equation is sufficient to describe the spatial
propagation of light in a pencil-shaped sample with a Fresnel number
${\cal F}= A / \lambda L$ near unity ($A$ is the cross-sectional area
of the sample and $\lambda$ is the optical wavelength) \cite{Raym85}.

If the strong field is not depleted in the process of quantum field
absorption and if most of the atomic population stays in the initial
level $1$, Eqs.(\ref{Bloch}-\ref{Max}) can be integrated to get
\begin{mathletters}
\begin{eqnarray}
\hat{Q}(z,\tau ) &=&e^{-\Gamma \tau }\hat{Q}(z,0)-e^{-\Gamma \tau
}\int_{0}^{z}dz^{\prime }\,\hat{Q}(z^{\prime },0)\sqrt{\frac{a(\tau )}
{z-z^{\prime }}}J_{1}(2\sqrt{a(\tau )(z-z^{\prime })})  \nonumber \\
&-&i\kappa _{1}\int_{0}^{\tau }d\tau ^{\prime }\,e^{-\Gamma (\tau -\tau
^{\prime })}\hat{E}_{q}(0,\tau ^{\prime })E_{s}(\tau ^{\prime })J_{0}(2
\sqrt{z(a(\tau )-a(\tau ^{\prime }))})+\int_{0}^{\tau }d\tau ^{\prime
}\,e^{-\Gamma (\tau -\tau ^{\prime })}\hat{F}(z,\tau ^{\prime })  \nonumber
\\
&-&\int_{0}^{\tau }d\tau ^{\prime }\,\int_{0}^{z}dz^{\prime }\,e^{-\Gamma
(\tau -\tau ^{\prime })}\hat{F}(z^{\prime },\tau ^{\prime })
\sqrt{\frac{a(\tau )-a(\tau ^{\prime })}{z-z^{\prime }}}
J_{1}(2\sqrt{(a(\tau )-a(\tau
^{\prime }))(z-z^{\prime })})  \label{Q} \\
\hat{E}_{q}(z,\tau ) &=&\hat{E}_{q}(0,\tau )-i\kappa _{2}E_{s}(\tau
)e^{-\Gamma \tau }\int_{0}^{z}dz^{\prime }\,\hat{Q}(z^{\prime },0)J_{0}(2
\sqrt{a(\tau )(z-z^{\prime })})  \nonumber \\
&-&\kappa _{1}^{\ast }\kappa _{2}E_{s}(\tau )\int_{0}^{\tau }d\tau ^{\prime
}\,e^{-\Gamma (\tau -\tau ^{\prime })}\hat{E}_{q}(0,\tau ^{\prime
})E_{s}^{\ast }(\tau ^{\prime })\sqrt{\frac{z}{a(\tau )-a(\tau ^{\prime })}}
J_{1}(2\sqrt{z(a(\tau )-a(\tau ^{\prime }))})  \nonumber \\
&-&i\kappa _{2}E_{s}(\tau )\int_{0}^{\tau }d\tau ^{\prime
}\,\int_{0}^{z}dz^{\prime }\,e^{-\Gamma (\tau -\tau ^{\prime })}\hat{F}
(z^{\prime },\tau ^{\prime })J_{0}(2\sqrt{(a(\tau )-a(\tau ^{\prime
}))(z-z^{\prime })})  \label{E}
\end{eqnarray}
\end{mathletters}
where $\tau =t-z/c$, and $a(\tau )=\kappa _{1}^{\ast }\kappa
_{2}\int_{0}^{\tau }d\tau ^{\prime \prime }\,|E_{s}(\tau ^{\prime
\prime })|^{2}$ and $\hat{Q}(z,0)$ is the initial atomic coherence.

Integrating Eq.(\ref{Q}) over space we obtain the collective atomic
spin operator, which is the atomic variable on which the quantum light
field is mapped.
\begin{eqnarray}
\hat{{\cal Q}}_{L}(\tau ) \equiv n\int_{0}^{L}\,dz\,\hat{Q}(z,\tau ) 
&=&ne^{-\Gamma \tau }\int_{0}^{L}\,dz^{\prime }J_{0}(2\sqrt{a(\tau
)(L-z^{\prime })})\hat{Q}(z^{\prime },0)  \nonumber \\
&+&n\int_{0}^{\tau }\,d\tau ^{\prime }e^{-\Gamma (\tau -\tau ^{\prime
})}\int_{0}^{L}\,dz^{\prime }J_{0}(2\sqrt{(a(\tau )-a(\tau ^{\prime
}))(L-z^{\prime })})\hat{F}(z^{\prime },\tau ^{\prime })  \nonumber \\
&-&in\kappa _{1}\int_{0}^{\tau }\,d\tau ^{\prime }e^{-\Gamma (\tau -\tau
^{\prime })}\hat{E}_{q}(\tau ^{\prime })E_{s}(\tau ^{\prime })\sqrt{\frac{L}{
a(\tau )-a(\tau ^{\prime })}}J_{1}(2\sqrt{a(\tau )-a(\tau ^{\prime })L})
\label{Q-L}
\end{eqnarray}
Eq.(\ref{Q-L}) is the main result of this Letter. The first term
represents the decaying memory of the initial atomic coherence in the
sample, the second term is the contribution from the Langevin noise
associated with the decay of the coherence, and the last term
represents the contribution from the absorbed quantum light. It is
thus the last term, that describes the quantum memory capability of
the atomic system. Note that the strong classical field
$E_{s}(\tau^{\prime})$ can be turned on and off, so that only the
value of the quantum field in a certain time window is mapped onto the
atomic system, where it is subsequently kept. We assume that the rate
$\Gamma $ is dominated by the power broadening contribution
$\Gamma_{s}$ when the classical field is turned on, and it can be
quite small $\Gamma =\Gamma_{0}$ when the classical field is turned
off to ensure long storage times.  If the quantum field pulse
$\hat{E}_{q}(\tau)$ and the overlapping classical pulse $E_{s}(\tau)$
are long enough so that $\Gamma \tau \gg 1$ the initial atomic state
decays and the state determined by $\hat{E}_{q}(\tau )$ emerges
instead. After the light pulses are turned off, the atomic ''memory''
state decays slowly with the rate $\Gamma_{0}$.

As an example of storing a quantum feature of light in atoms let us
consider storing a squeezed state, which plays an important role in
quantum information with continious variables \cite{Brau98b}.  For
infinitely broadband squeezed light the quadrature operator
$\hat{X}_{q}(z,\tau) = \text{Re} \hat{E}_{q} (z,\tau )$ on the entry
face of the sample can be written as $\langle \hat{X}_{q}(0,\tau)
\hat{X}_{q}(0,\tau^{\prime }) \rangle = 2 \pi \hbar \omega / c \langle
X_{0}^{2}\rangle \delta (\tau -\tau^{\prime })$, where $\langle
X_{0}^{2}\rangle$ is the dimensionless light noise, $\langle
X_{0}^{2}\rangle =1$ in the case of broad band vacuum. In steady-state
the variance of the atomic noise $\hat{X}=\text{Re}\hat{{\cal Q}}_{L}$
becomes
\begin{eqnarray}
\langle X^{2}\rangle &=& nL \left( e^{-\alpha} \left( I_{0}(\alpha)
+I_{1}(\alpha) \right) \right)  \nonumber \\
&+& nL \langle X_{0}^{2}\rangle \left( 1-e^{-\alpha} \left(
I_{0}(\alpha)+I_{1}(\alpha) \right) \right)
\label{var}
\end{eqnarray}
where $\alpha=aL/\Gamma$ is the optical depth of the sample,
$a=\kappa_{1}^{\ast} \kappa_{2} |E_{s}|^{2}$ and $I_{0}$ and $I_{1}$
are Bessel functions of the first kind. In the case of vacuum incident
on the sample we recover the atomic vacuum noise $\langle X^{2}\rangle
=nL$, the number of atoms per unit area. The second term in
(\ref{var}), represents the light contribution to atomic noise, it is
reduced when the light is squeezed, and in the case of ideally
squeezed light $\langle X_{0}^{2}\rangle =0$ only the first term
contributes to the atomic noise variance. We define the dimensionless
expression in the parenthesis as a mapping efficiency for the Gaussian
fields $\eta = (1-\langle X^2 \rangle / nL)/(1-\langle X_0^2 \rangle)$
(for ideally squeezed light $1-\eta$ quantifies the amount of spin
squeezing).  The results are plotted in Fig.\ref{fig:var} (solid line)
as a function of the optical depth $\alpha$.  Storing squeezing in
atoms with an efficiency higher than 90\% requires an atomic sample
with an optical depth of the order of $\gtrsim 60$. Note that by
absorption of EPR beams in separate atomic samples, we may, e.g.,
prepare entangled atomic gases, see also \cite{Polz99}.  If $\Gamma
\approx \Gamma_{s}$, and the decoherence is dominated by the strong
field that is required for the operational memory, then $\alpha \simeq
(3/2\pi) \lambda^{2} nL$, i.e., the optical depth is the same as for a
resonant narrowband field. The dependence on the optical depth arises
because the more squeezed light is absorbed in the sample, the more
the atoms become squeezed. If only a fraction of the light field is
absorbed, the atomic spins will not only be correlated with each other
but also with the field leaving the sample, and thus the squeezing
will be degraded, see also \cite{Molm99}.

Various schemes for quantum state exchange between light and atoms
based on cavity QED Raman-type interactions have been proposed in the
past \cite{Cira97,Park99,Zeng}. Quantum memory with a
microwave cavity field as storage medium has been demonstrated in
\cite{Mait97}.  The fact, that the present proposal does not utilize
high finesse cavities significantly simplifies the experimental
realization. The above result can be compared with the proposal
\cite{Kuzm97} and its experimental verification \cite{Hald99} for
squeezing the collective spin of an optically thick sample of $V$-type
excited atoms via the interaction with squeezed light. As opposed to
the theoretical bound of 50\% mapping efficiency found in
\cite{Kuzm97} the present proposal offers in principle a perfect
transfer of the state of light onto atoms.

A steady state analysis in frequency domain similar to that in
\cite{Kuzm97} leads to the following expression for the spectral
collective atomic spin operator
\begin{eqnarray}
\tilde{{\cal Q}}_{L}(\Delta ) &=&-\frac{i n}{\kappa _{2}E_{s}}\left(
1-e^{ik(\Delta )L}\right) \tilde{E}_{q}(\Delta )  \nonumber \\
&+&\int_{0}^{L}\,dz\frac{n}{\Gamma -i\Delta }e^{ik(\Delta) (L-z)} 
\tilde{F}(z,\Delta )  \label{spectra}
\end{eqnarray}
where $\Delta$ is the detuning from the two-photon resonance and
$k(\Delta)$ is the Lorentzian absorption profile $ik(\Delta) =
-a/(\Gamma -i\Delta )$.  The atomic noise variance $\langle
X^{2}\rangle =\int d\Delta \tilde{X} (\Delta )\tilde{X}(-\Delta )$
gives the same result as Eq.(\ref{var}).

The simplest approach to quantum field propagation in a medium is the
model of scattering by a collection of frequency-dependent beam
splitters \cite{Jeff}. Each beam splitter removes a small
fraction of a propagating light beam and it simultaneously couples in
a small fraction of vacuum into the beam. The result for the noise
spectrum of the transmitted light in our model coincides with such a
simplified treatment and is given by
\begin{equation}
\langle \tilde{X}^{2}(\Delta ) \rangle = \langle \tilde{X}_{0}^{2}(\Delta)
\rangle 
e^{-\frac{a\Gamma L}{\Gamma ^{2}+\Delta^{2}}} + \left( 1-e^{
-\frac{a\Gamma L}{\Gamma^{2} + \Delta ^{2}}} \right) 
\end{equation}
For infinite bandwidth squeezed incident light this spectrum
approaches the vacuum value $1$, for the frequencies where light is
strongly attenuated.  The width of this noise region grows with
optical depth of the system. It is within this spectral region that
quantum features of the light field are transferred onto atoms.

In the case of the finite bandwidth of ideal squeezing
\cite{Shap} $\langle \hat{X}_{q}(0,\tau)
\hat{X}_{q}(0,\tau^{\prime }) \rangle \simeq 2 \pi \hbar \omega / c
(\delta (\tau -\tau^{\prime})- \Gamma_{q} / 2e^{-\Gamma_{q} | \tau -
\tau^{\prime }|})$, calculations based on either Eq.(\ref{Q-L}) or
Eq.(\ref{spectra}) have to be carried out numerically and the mapping
efficiencies for different spectral widths of squeezing $\Gamma_{q}$
are shown in Fig.\ref{fig:var}. We observe in the figure that when the
entire bandwidth of squeezed light is completely absorbed in the
sample, further growth of the optical depth leads only to the
reduction of the spin squeezing, because the atoms which are not
reached by the squeezed light are subject to the standard vacuum
noise.

The macroscopic number of atoms in our atomic sample, of which most
remain in the ground state, allows us to replace the sum of fermionic
atomic operators by an effective bosonic operator $\hat{{\cal Q}}_{L}$
matching the bosonic operator of the light field. This restriction
should be kept in mind when comparing our results to other analyses of
spin-squeezing \cite{Wine}.

A suitable experimental setup for realization of the storage of field
correlations in atoms is the cold atom fountain, e.g. as used in a
frequency standard. A recent paper \cite{Sant99} reports operation of
a laser cooled cesium fountain clock in the quantum limited regime
meaning that the variance $\langle X^{2}\rangle =nL$ of the collective
atomic spin associated with the $F=4,m=0$ -- $F=3,m=0$ two level
system has been achieved. This means that the setup is suitable for
the observation of squeezing of $\langle X^{2} \rangle$. The
decoherence time $\Gamma_{0}$ of the order of a second reached in the
atomic standard setup in principle allows quantum memory on this time
scale. We thus propose to prepare atoms in the $F=3,m=0$ state (our
state $1$, the level $F=4,m=0$ plays the role of our state $3$) and to
illuminate them by a Raman pulse containing the squeezed vacuum and
the strong field as described above. After the pulse and after some
delay the atoms are interrogated in a microwave cavity where their
collective spin state is analyzed to verify that the memory works.

We now wish to address the experimental requirements for our
proposal. For our two-level analysis to be valid, we assume that
$\Delta \gg \Gamma_{q}, \Gamma_{s}, \gamma_{i}$ and $\sigma_{R} \gg
\sigma_{\text{2-level}}$ where $\sigma_{R}=(6\pi)^{4} c^{8}
I_{\text{sat}}^{2} / (2 \Gamma_{q} S \omega^{11} \hbar^{3}
\Delta_{i}^{2})$ is the stimulated Raman cross section for the quantum
field, $S=I_{s}/I_{\text{sat}}$ is the saturation parameter and
$I_{\text{sat}}=\omega^{6}/ (9\pi c^{5}) \sum \mu_{1i} \mu_{3i}$ is
the saturation intensity for the strong field for $1 \leftrightarrow
i$, $3 \leftrightarrow i$ transitions, $\sigma_{\text{2-level}} = 3
\lambda^{2} \gamma_{i}^{2} / (8\pi \Delta_{i}^{2})$ is the spontaneous
2-level cross section. In order to carry out the steady state solution
of (\ref{Bloch}-\ref {Max}) we assume $\Gamma_{s} \gg
\tau_{\text{pulse}}^{-1}$ - where $\tau_{\text{pulse}}$ is the
duration of the Raman pulse. Finally, the condition on the bandwidth
of the quantum field $\Gamma_{q} \gg \tau_{\text{pulse}}^{-1}$ ensures
that the pulse is long enough to contain all relevant correlations of
the quantum state of the field. It is possible to satisfy all those
conditions with the following set of parameters:
$\Gamma_{q}=10^{7}$Hz, $\Delta = 10^{9}$Hz, $S>4$,
$\tau_{\text{pulse}} = 10$ $m$sec. With the resonant optical depth of
$20$ achievable for $5 \times 10^{5}$ atoms a mapping efficiency
exceeding 80\% is possible (Fig.\ref{fig:var}). After the pulse is
switched off the memory time $\Gamma_{0}^{-1}$ is set by the free
evolution of the $F=4,m=0$ -- $F=3,m=0$ system and as mentioned above
it can be as long as a second.

We have analyzed the possibility to transfer (write-down) a quantum
state of light onto an atomic sample. And we have suggested how to
perform a delayed measurement of the quantum state.  We will now
briefly discuss how to map the atomic state back onto a light field by
interspecies teleportation \cite{Maie98}. To realize an effective
teleportation of an atomic collective spin onto a light beam we
suggest an approach similar to teleportation of light
\cite{Furu98,Brau98,Vaid94} with EPR correlated light beams and a
beam-splitter type interaction between one of the beams and the atomic
collective spin. Making a homodyne measurement of the light quadrature
and a Ramsey measurement of the atomic spin we may employ the protocol
used for light teleportation \cite{Furu98,Brau98,Vaid94} and restore
the atomic state in the other light beam.

To realize the ``beam-splitter'' we send a short pulse
($\tau_{\text{pulse}} \Gamma \ll 1$ - so that dissipation processes
do not take place) of one of the EPR beams through our atomic sample
in the small optical depth regime ($\alpha \rightarrow a
\tau_{\text{pulse}} L \ll 1$). In our scheme the switching from high to
small optical depth is made simply by adjusting the intensity of the
coupling field $E_{s}$. In the weak coupling regime (small optical
depth) the interaction between light and atoms (\ref{E}) - (\ref{Q-L})
can be described by a linear approximation leading to a
``beam-splitter'' type interaction. Introducing a new rescaled atomic
operator $\hat{q} = (nL)^{-1/2} \hat{{\cal Q}}_{L}$ and the field
``area'' operator $\hat{\theta} = \sqrt{\lambda / 2 \pi \hbar
\tau_{\text{pulse}}} \int_{0}^{\tau_{\text{pulse} }} d \tau^{\prime}
\hat{E}_{q}(\tau^{\prime})$ we obtain:
\begin{mathletters}
\begin{eqnarray}
\hat{q}_{\text{out}} &=&\hat{q}_{\text{in}}-ir\hat{\theta}_{\text{in}} \\
\hat{\theta}_{\text{out}} &=&\hat{\theta}_{\text{in}}-ir\hat{q}_{\text{in}}
\end{eqnarray}
\end{mathletters}
The condition for such a linearization is a weak interaction, hence
our ``beam-splitter'' is highly asymmetric, $r = \sqrt{
\alpha} = \sqrt{\sigma_{R} \tau_{\text{pulse}} L /
\Gamma_{q}} \ll 1$. Teleportation with asymmetric beam splitters is
possible but it requires a higher degree of correlation in the EPR
beams. A simple estimate suggests, that the residual noise in the EPR
pair must be smaller than $r$. 
If one assumes a stronger coupling in order to approach the symmetric
beam-splitter case, the field probes a component of the atomic
coherence, which deviates from the uniform integral in Eq.(\ref{Q-L})
due to the spatial variation of the probe light. If, for example, the
probe is damped by a factor of order 2, it is reasonable to decompose
the probed atomic coherence as a roughly even mixture of the uniform
integral $\hat{{\cal Q}}_{L}$ and a ``noise'' operator which we, for
simplicity may assume to be the standard vacuum noise. This noise is
comparable to the ``quduty'' of noise \cite{Brau98} of a direct
detection of the atomic ensemble and reconstruction of a corresponding
field state (``classical teleportation'').

We are grateful to Prof.\ Sam Braunstein for stimulating discussions
of the atomic teleportation. This research has been funded by the
Danish Research Council and by the Thomas B.\ Thriges Center for
Quantum Information. AK acknowledge support of the ESF-QIT programme.


\end{document}